\documentclass[fleqn,usenatbib]{mnras}

\usepackage{newtxtext,newtxmath}

\usepackage[T1]{fontenc}
\usepackage{ae,aecompl}

\usepackage{todonotes}

\usepackage{graphicx}	
\usepackage{amsmath}	
\usepackage{amssymb}	

\title[Impact of Helium on Multiple Population Dynamics]{The Effect of Stellar Helium Abundance on Dynamics of Multiple Populations in Globular Clusters}

\author[Fare, Webb, \& Sills]{
Amy Fare,$^{1,3}$\thanks{E-mail: ifare2@uwo.ca; asills@mcmaster.ca}
Jeremy J. Webb,$^{2}$
Alison Sills$^{1}$
\\
$^{1}$Department of Physics \& Astronomy, McMaster University, 1280 Main Street West, Hamilton, ON, L8S 4M1 CANADA\\
$^{2}$Department of Astronomy \& Astrophysics, University of Toronto, 50 St George St, Toronto, ON, M5S 3H4 CANADA\\
$^{3}$Department of Physics \& Astronomy, Western University, 1151 Richmond St, London, ON, N6A 3K7 CANADA\\
}

\date{Accepted 2018 August 29. Received 2018 July 31; in original form 2018 June 5}

\pubyear{2018}

\begin{document}
\label{firstpage}
\pagerange{\pageref{firstpage}--\pageref{lastpage}}
\maketitle

\begin{abstract}

We incorporate a semi-analytic formula for the main sequence lifetime of helium-rich stars in \textit{N}-body simulations of multiple population globular clusters to investigate how the enriched helium stars impact the dynamics of globular clusters. We show that a globular cluster with a helium-rich concentrated population will be slightly smaller than a globular cluster with a normal-helium second generation, with the largest difference seen in the extended normal-helium population. This effect is shown both for a cluster in isolation and one in a realistic Milky Way tidal field. We show that this effect is a result of mass segregation, and the earlier loss of more massive stars in a helium-rich concentrated population due to their decreased main sequence lifetime. The two populations will therefore become dynamically mixed at a slightly earlier time than if they have the same helium abundance. Furthermore, we find that it is possible for the helium-enriched population to become more extended than the normal-helium population if it forms with a low initial concentration or the cluster is able to evolve for a large number of relaxation times. We conclude that the dynamical effects of helium on stellar masses are modest, and that the initial concentration of the two populations and the strength of the Milky Way tidal field are more important in determining the relative radial distributions of multiple populations.
\end{abstract}



\section{Introduction}

Globular clusters were, for many years, considered to be the canonical example of a simple stellar population, but are now understood to be more complex. They contain more than one stellar population, differing in their abundances of, and with anti-correlations in, light elements such as carbon and nitrogen, sodium and oxygen, and aluminum and magnesium. While the astrophysical source of these elemental abundance patterns is still unclear, the general picture is that some source of hot hydrogen burning was present in or near the early cluster, and that material was made available to form the stars that make up the anomalous population(s). For a more complete discussion of the observational evidence and some of the proposed mechanisms, see the recent review by \citet{bastianlardo2017}.

One implication of this picture is that the anomalous population must also be enriched in helium. Indeed, there is both indirect \citep[e.g.][]{bedin__2004} and direct \citep[e.g.][]{dupree2011} evidence of a helium spread. The amount of helium required to match the observations varies from cluster to cluster. Claims of an extreme population with Y$\sim$0.4 ($\Delta Y = 0.15$) in NGC 2808 is the highest mentioned in the literature \citep{piotto2007triple}, while in M3 we see only a very modest helium increase of $\Delta Y = 0.01-0.02$ \citep{valcarceM3helium}. Typical maximum values are of order of $\Delta Y \approx 0.04 - 0.1$ \citep{miloneHB2014}, corresponding to Y $\approx$ 0.28 - 0.34. 

Stars enhanced in helium are bluer and brighter than their normal-helium counterparts of the same mass, and have shorter main sequence lifetimes. Therefore, there will be a difference in the mass of stars at the turnoff of a globular cluster at a given age -- start with normal helium will be more massive than those with higher helium. We know that the helium-rich population makes up around 50\% of the total mass of the cluster \citep{bastianlardo2017}. Therefore, if the helium enhancement is sufficiently large, the reduced mass of the main sequence stars and the accelerated evolution of stars into remnants could change the dynamical evolution of the cluster. 

The dynamics of multiple-population globular clusters has been investigated by a number of groups. Of particular interest is the spatial distribution of the helium-rich population stars in the cluster. Some of the scenarios for the formation of multiple populations involve a central concentration of polluted gas which then forms the helium-rich stars, and so evidence of a central concentration of those stars may pinpoint the formation mechanism. Indeed, analyses of red giant stars in globular clusters have found enriched populations, as distinguished by their high Na abundance, to be more centrally concentrated than their normal-abundance counterparts \citep{carretta2009anticorrelation,carretta2010radial}. The globular cluster NGC 6362, however, has been determined to have spatially mixed populations \citep{dalessandro2014first}. 

The central concentration of helium-enhanced stars has been modified from its initial distribution by the 10-12 Gyr of dynamical evolution since the two populations were formed. \cite{vesperini2013dynamical} studied the time-scales of spatial mixing using \textit{N}-body simulations. They found that the cluster takes longer to spatially mix if  their ``SG" (enriched) population is initially more centrally concentrated. They also found that mixing occurs more quickly for a cluster in a tidal field than a cluster in isolation. \cite{decressin2008evolution} suggested that globular clusters should erase any initial central concentration of the enriched population within a few relaxation times if gas expulsion is also taken into account. However, it has been found that a kinematic signature of the initial central concentration may remain in form of differential rotation between the two populations \citep{henaultbrunet15}.

\cite{miholics2015dynamics} modelled systems in a more realistic Milky Way tidal field. They found that the high mass loss rate experienced by a cluster with the orbit of globular cluster NGC 6362 in a Milky Way tidal field accelerates its spatial mixing. Normal-abundance stars in the outer regions of the cluster are preferentially stripped from the cluster, causing that population to shrink at a faster rate than the originally centrally concentrated helium-rich population. In general, clusters are spatially mixed when they have lost 70-80\% of their initial mass. 

Recently it was reported that in the dynamically evolved cluster M80, the normal-helium population is more centrally concentrated that the helium-rich population\citep{dalessandroM80}. Using \textit{N}-body simulations, they argue that this peculiar pattern does not depend on the initial relative concentration of the two populations. Instead, they propose that the red giants in the helium-rich population are, on average, 0.05 to 0.1 $M_{\odot}$ less massive than those in the normal-helium population.  This mass difference is consistent with a difference in helium of about $\Delta Y = 0.05$.

In this paper, we investigate the dynamical effects of different helium abundances in two populations. We refer to the normal helium population as P1 and the helium-rich population as P2 but we remind the reader that the numbers 1 and 2 do not necessarily indicate an order in which the stars formed. In section 2, we discuss our modifications to the \textit{N}-body code NBODY6 \citep{aarseth03_nbody6} to allow us to follow two populations with different stellar lifetimes and describe the simulations we performed. Our results are discussed in section 3, and we summarize in section 4.

\section{Methods}

\subsection{Simulating helium-rich stellar populations}

Stellar evolution in $N$-body simulations is typically handled through functional fitting to stellar evolutionary tracks. NBODY6 uses the {\sc SSE} package \citep{hurley_comprehensive_2000}, which calculates stellar quantities such as lifetimes, luminosity, and temperature as a function of time, stellar mass, and metallicity. The default assumption is that the helium abundance varies with metallicity as $\Delta Y/ \Delta Z = 2.0$. If we wish to evolve two populations with the same metallicity but different helium abundances, we must modify the stellar evolution formulae appropriately.

A series of evolutionary tracks for helium values from Y=0.248 to Y=0.8 were calculated by \cite{chantereau_evolution_2015}, for masses appropriate for the present day population of main sequence stars in globular clusters. Using these tracks, we determined that the ratio of main sequence lifetimes depends strongly on helium abundance and very weakly on stellar mass. We found that the ratio of helium-rich to normal-helium main sequence lifetimes can be given by:
\begin{equation}
t/t_{Y=0.248}=10^{-1.074(Y-0.248)^2-2.776(Y-0.248)}
\label{tms}
\end{equation}

\begin{figure}
    \includegraphics[width=\columnwidth]{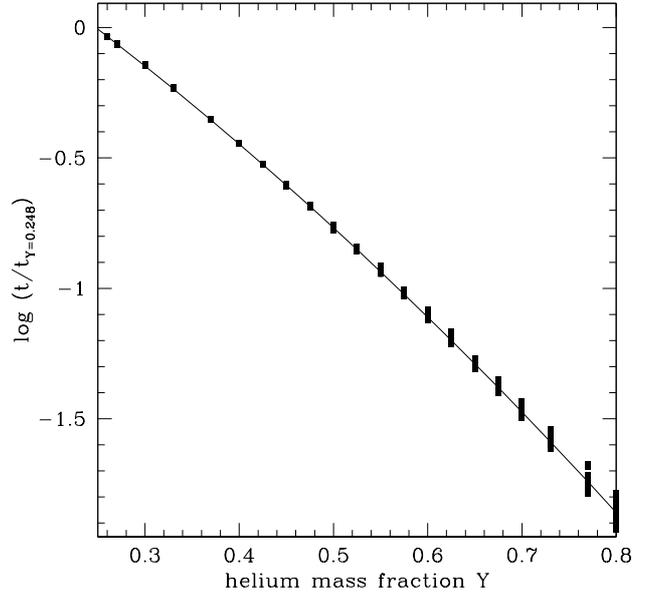}
    \caption{Logarithm of main sequence lifetime ratio (lifetime compared to a star of the same mass with Y=0.248) vs helium abundance. The solid points show the ratio for masses between 0.3 and 0.8 $M_{\odot}$ from the models of \citet{chantereau_evolution_2015}, and the line gives our fit as described in the text.
   \label{fig:heliumfit}}
\end{figure}

\noindent where $t$ is the main sequence lifetime for a star with a helium abundance of $Y$, and $t_{Y=0.248}$ is the main sequence lifetime of the lowest Y value from the \citet{chantereau_evolution_2015} evolutionary tracks. The main sequence lifetime ratios from the tracks for all masses, and our fit, are shown in Figure \ref{fig:heliumfit}. At the highest values of Y, we see a spread in the lifetime ratio for stars of different mass, which is not apparent at the lowest helium abundances. The highest inferred helium abundances in a globular cluster is about Y=0.4 so having a mass-independent fit is appropriate for the question we are considering.  We confirmed the validity of the fit with the helium-rich evolutionary tracks of \citet{dotter2008dartmouth}, which cover a smaller range in helium but a larger range in mass. Therefore, we use this functional form to modify the {\sc SSE} main sequence lifetimes for our high-helium population for stars of all masses.

\subsection{Implementation of high helium stellar evolution in {\sc NBODY6}}

In order to model star clusters made up of two sub-populations with different properties, several modifications were made to the existing {\sc NBODY6} code. Our modified code, named {\sc NBODY6MP} \footnote{https://github.com/webbjj/nbody6mp.git}, builds off an already modified version for {\sc NBODY6} which allows for clusters to be modelled in arbitrary tidal fields \citep{renaud11_nbody6tt} and makes three important additions. The first addition is to tag stars by population number (e.g. P1 and P2), so stars in different sub-populations can be tracked in real-time. The second addition is to allow for stars in different sub-populations to have different metallicities and helium abundances. Finally, when the helium abundance of a sub-population is not simply 0.24+2Z (the default Y value in NBODY6), we instead calculate the main-sequence lifetime of stars in the sub-population using Equation \ref{tms}. All other properties of stars with non-standard Y values retain their default values (i.e. as if they still have helium abundances equal to Y=0.24+2Z). 

\subsection{Initial Conditions}

To create clusters with two populations, we follow the method outlined in \citet{miholics2015dynamics}. First, we use  McLuster \citep{kupper2011mass} to generate 20 000 P1 stars, distributed following a King density profile with $W_0=7$ \citep{king1966structure} with an initial half-mass radius of 5.0 pc, in virial equilibrium, with a Kroupa mass function \citep{kroupa2001variation} between 0.1 and $50M_\odot$. We then scale the same distribution to half its radius to act as the second population (P2). We combine the two distributions and adjust the stellar velocities so that the ensemble is in virial equilibrium. As a result, we have a cluster with 40 000 stars that contains a centrally concentrated second population. Our choices of a 50:50 split between the two populations, as well as the particular initial central concentration for the helium-rich population, are guided by general constraints from observations of present-day clusters. At the same time, these choices should be sufficiently extreme that any effects of helium enhancement should be clearly visible.

We ran simulations in which the P2 stars were assigned the same helium abundance as the P1 stars (Y=0.24 + 2Z = 0.2402 for our chosen Z=0.0001), as a benchmark. We ran a second set of simulations in which we assigned a helium abundance of Y=0.32 to our P2 stars. To test the effect of helium on the evolution of the cluster, our value of Y needed to be large enough for there to be a substantial difference in the lifetimes of the two populations, but not so large that it was out of line with the observational values from globular clusters. We also evolved these systems in two tidal environments -- an isolated system in which no tidal stripping occurred, and on a circular orbit at 20 kpc in the Milky Way potential described in \citet{miholics2015dynamics}. We ran our simulations for at least 10 Gyr. 

\section{Results}
\subsection{Stellar evolution with a helium-rich population}

\label{stellar_evolution}

\begin{figure}
    \centering
    \includegraphics[width=\columnwidth]{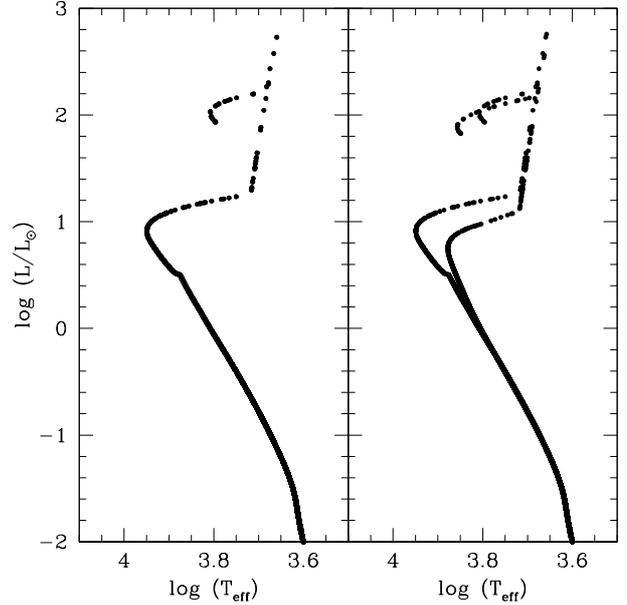}
    \caption{HR diagram of a cluster with two normal-helium populations (left), and a cluster with both a normal-helium and helium-rich population (right) at 4 Gyr. The double sequences seen in the right panel are caused by the difference in main sequence lifetimes of the two populations.}
   \label{hr_4000Myr}
\end{figure}

Figure \ref{hr_4000Myr} shows Hertzsprung-Russell diagrams of two clusters at 4 Gyr, with temperatures and luminosities taken from the {\sc SSE} implementation in {\sc NBODY6MP}. On the left is the diagram taken from our simulation in which both populations have the same helium abundance. On the right is the result when we give the P2 population Y=0.32. We see the two turnoffs and horizontal branches, caused by the modified main sequence lifetime imposed on the helium-rich population. We note that there is no spread on the lower main sequence or red giant branch, as we have not modified any stellar properties (e.g. temperature, luminosity) other than main sequence lifetime. Our emphasis for this work is to understand the implications of the change in mass on the stellar dynamics caused by the shorter lifetimes of the helium-rich population. 

We calculated the mass difference between the stars at the turnoff for the two different populations as a function of time. After the most massive stars evolve away, the mass difference is essentially constant at 0.15 $M_{\odot}$ for our difference in helium of $\Delta Y = 0.08$. However, this mass difference only affects a small fraction of stars in the clusters -- just those in the helium-rich population which have evolved off the main sequence earlier than their counterparts in the helium-normal population. We also calculated the average mass in each population for a simulation evolved without a tidal field. We found that the difference in average mass is constant at $\sim$0.01 $M_{\odot}$ after the most massive stars have become remnants. 

\subsection{Dynamical effects of a He-rich population}

\label{res_isol}

\begin{figure}
    \centering
    \includegraphics[width=0.5\textwidth]{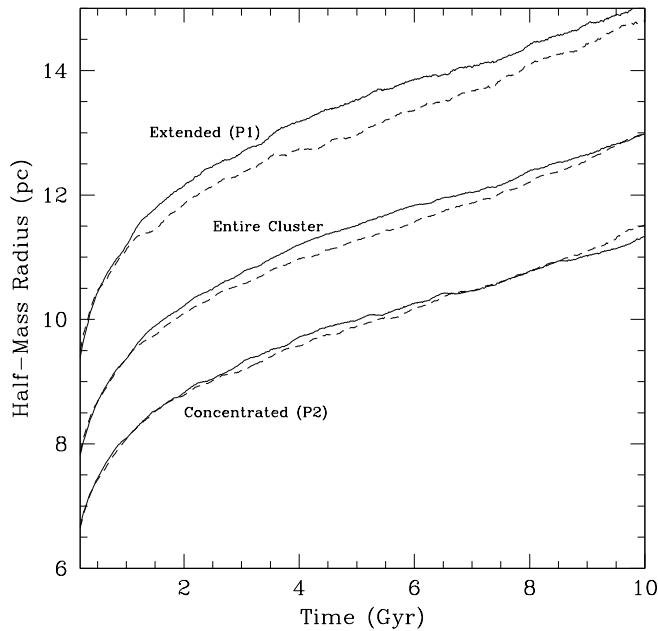}
    \caption{Half-mass radii of clusters and their sub-populations in isolation. Solid lines show the half-mass radii for the cluster in which both populations have the same normal helium abundance, while dashed lines show the half-mass radii for the simulation in which the more centrally concentrated cluster has higher helium. From top to bottom, the lines correspond to the half-mass radius of the more extended population (P1), that of the entire cluster, and that of the more concentrated population (P2).}
   \label{half_mass_isolation}
\end{figure}

First, we look at the effect of a helium-rich population on the size of a cluster in isolation. Figure \ref{half_mass_isolation} shows the half-mass radii of the entire cluster, as well as the half-mass radii of each population individually. From top to bottom, we show the half-mass radius of the P1 (extended) population, the entire cluster, and then the P2 (concentrated) population. The solid lines correspond to the simulation in which both the P1 and P2 populations have the same, primordial, helium abundance, while the dotted lines show the simulation in which the P2 concentrated population has Y=0.32 while the P1 extended population has Y=0.2402. After an initial rapid expansion of all populations caused by stellar-evolution driven mass loss, the size of the clusters gradually expand because there is no tidal truncation. After approximately 1 Gyr, the simulation containing a helium-rich population has a slightly smaller half-mass radius than the simulation in which both populations have the same helium abundance. The extended P1 population shows the largest difference, while the concentrated P2 populations are essentially the same.

\begin{figure}
    \centering
    \includegraphics[width=0.5\textwidth]{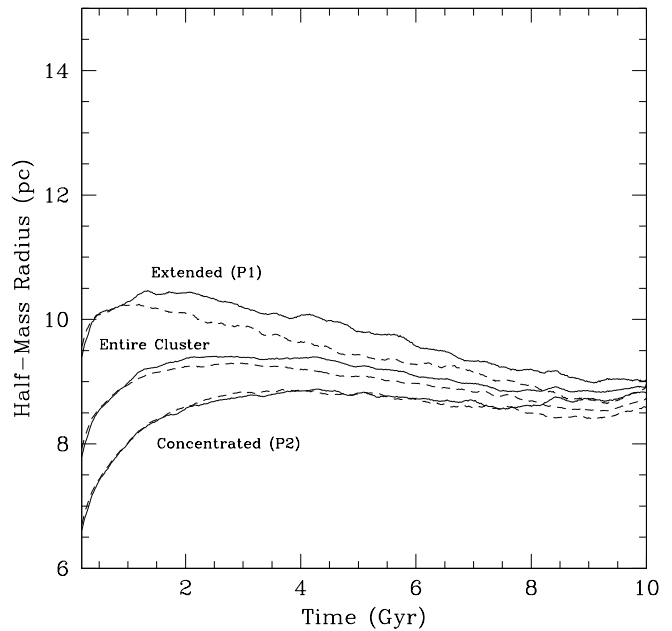}
    \caption{Half-mass radii as a function of time for a cluster evolving in a tidal field on an orbit at 20 kpc from the centre of the Milky Way. The line styles and orientation are the same as in Figure \ref{half_mass_isolation}}
   \label{half_mass_tidal}
\end{figure}

We see a similar behaviour for a cluster evolved in the tidal field of the Milky Way. Figure \ref{half_mass_tidal} shows the same clusters placed on a circular orbit at 20 kpc from the centre of the Galaxy. As expected \citep{miholics2015dynamics}, the cluster only expands for the first 2 Gyr or so, and then the Galactic tidal field removes stars from beyond the tidal radius, effectively reducing the total mass and hence the half-mass radius. After about 8 Gyr, the P1 and P2 populations are essentially mixed in both simulations. The simulation with the helium-rich concentrated P2 population (dashed lines) also becomes smaller more quickly than the simulation with two helium-normal populations, and possibly shows signs of mixing more quickly although the difference is small. 

In both simulations, the extended P1 population shows the largest difference caused by the change in helium abundance. On the face of it, this is surprising as the two P1 populations have the same helium abundances and have the same initial configuration (in terms of stellar masses, positions, velocities, etc). We would expect that these two populations would evolve in a similar way, while the two P2 concentrated populations should evolve differently. However, this neglects the effects of the interaction between the two populations. 

The primary mode of interaction between the P1 and P2 populations is through mass segregation, the tendency for the most massive stars to sink to the centre of the cluster. Because of the difference in lifetimes, the most massive stars in a normal helium population correspond to stars in a helium-rich population which have have already left the main sequence, lost some mass, and become white dwarfs. Not only does the simulation with the helium-rich population lose more mass in total during this process, but mass segregation will occur differently. In our cluster with a helium-rich component, we have lost half of the stars which would be at the turnoff in the normal helium case. Therefore, other than stellar remnants, the most massive stars in the cluster are entirely from the P1 population, and they segregate much faster than their counterparts in the simulation in which both populations have the same helium abundance. As a result, the half-mass radius is significantly smaller for the extended P1 population. 

\begin{figure}
    \includegraphics[width=\columnwidth]{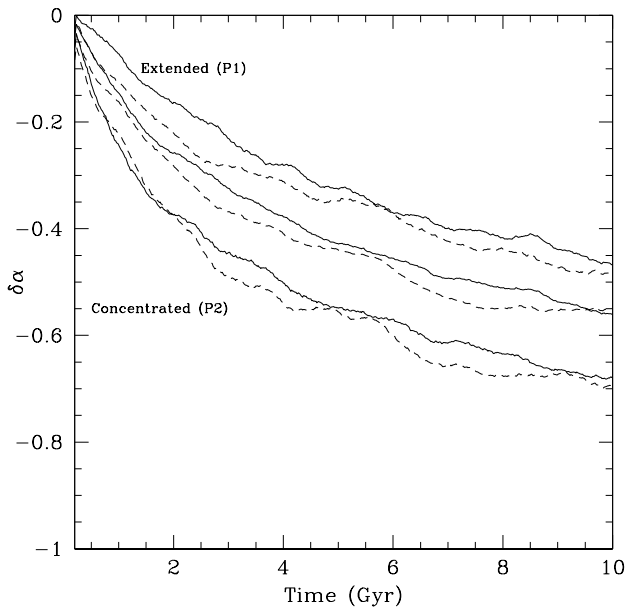}
    \caption{Change in mass function slope over the cluster as a function of time, for the simulations shown in figure \ref{half_mass_isolation}. The line styles and orientations are as in that figure. 
   \label{fig:massseg}}
\end{figure}

To test this explanation, we measure the mass segregation for the cluster as a whole, as well as each population separately, using the method outlined in \citet{webb_deltaalpha}. Figure \ref{fig:massseg} illustrates how the the slope of the stellar mass function $\alpha$ (for stars between 0.3 and 0.8) varies with clustercentric radius r as a function of time by calculating $\delta_\alpha = \frac{d \alpha}{d ln(\frac{r}{r_m})}$, where $r_m$ is the half-mass radius of the cluster. We calculate $\delta_\alpha$ for P1 and P2 stars separately, using the individual $r_m$ of each population, as well as for the cluster as a whole. The solid lines show the mass segregation evolution for the simulation in which both populations have normal helium, while the dotted lines show the segregation for the simulation in which the P2 population is helium-rich. As expected, the P1 population shows a significantly enhanced segregation signature early on when the P2 population is helium-rich, which persists through the simulation.

\begin{figure}
    \includegraphics[width=\columnwidth]{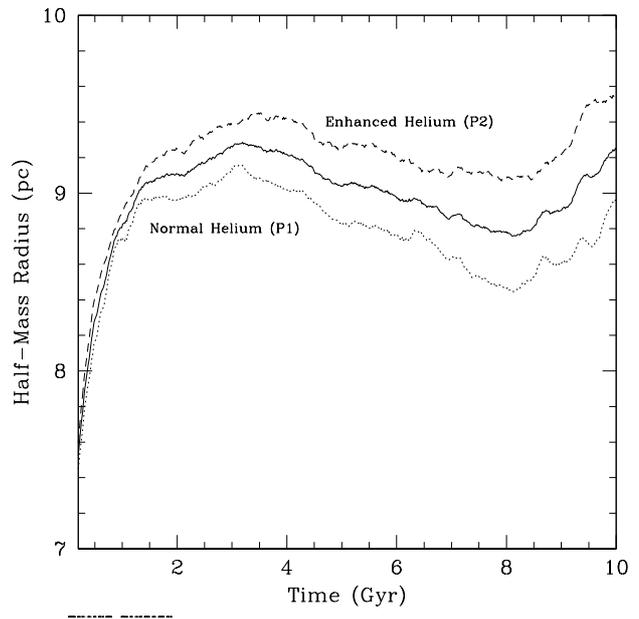}
    \caption{Half mass radii as a function of time for a simulation in which the two populations had the same initial concentration. The population with enhanced helium (P2, dashed line) becomes more extended than the population with normal helium (P1, dotted line). The solid line shows the half-mass radius of the entire cluster. 
   \label{fig:c1}}
\end{figure}

Finally, we ran one simulation where we expected to see the largest difference between an enhanced helium population and one with normal helium. This simulation is the same as the one shown in figure \ref{half_mass_tidal} with dashed lines, except that we allowed the two populations to have the same initial size -- in other words, the populations were initially fully mixed. Figure \ref{fig:c1} shows the half-mass radii as a function of time. The overall evolution of the cluster is shown as the solid line. The enhanced helium population becomes more extended than the normal-helium population, as expected. We also ran this simulation but with the same helium abundance for both populations. We saw no significant difference in the half-mass radii of the two populations or the cluster as a whole, confirming that the result shown here is caused by the change in mass and therefore extra mass segregation, combined with tidal stripping, caused by the helium abundance of the stars.

\section{Discussion \& Conclusions}

We investigated the dynamical evolution of globular clusters with multiple populations, including the effect of enhanced helium abundance on the main sequence lifetime of the helium-rich population. We find that the helium enrichment of an initially centrally-concentrated population has a small effect on the overall size of the cluster, making the whole system slightly smaller. The extended P1 population is smaller in the cluster with the helium-rich P2 population, meaning that spatial mixing between the two populations will occur slightly sooner than we have predicted using constant-helium models. Furthermore, once the two populations are spatially mixed, we find that the helium-rich P2 population is able to become more extended than the P1 population due to the effects of mass segregation and P2 stars having a lower mean mass.

Our results confirm the conclusion of \citet{dalessandroM80} that a stellar mass difference caused by the inferred helium spread could speed up the mixing of the two populations. However, the helium difference chosen for our simulations was fairly large, and yet the difference between the times of full mixing in our simulation is modest. We see larger dynamical effects on the relative sizes of the two populations caused by the presence and strength of the tidal field of the Milky Way, and also from the initial concentration of the helium-rich population. Hence, while a difference in He abundance is necessary, additional models have shown that a significantly less concentrated P2 is initially required in order for P2 stars to mix and then become more extended than P1 stars within 10 Gyr (as observed in M80). A similar effect may be observed in clusters with an initially more centrally concentrated P2 if they are able to evolve for a large number of relaxation times. 

Based on the comparison of the models presented in this paper with the investigations of central concentration and cluster orbit from \citet{vesperini2013dynamical} and \citet{miholics2015dynamics}, we conclude that the initial configuration of the two populations and the tidal environment of the cluster are more important for the dynamical evolution of the system than the helium abundances of the populations. We will explore these effects in more detail in a future paper.

\section*{Acknowledgements}
AIS and JJW are supported by NSERC. This work was made possible by the facilities of the Shared Hierarchical
Academic Research Computing Network (SHARCNET:www.sharcnet.ca) and Compute/Calcul Canada.



\bibliographystyle{mnras}
\bibliography{bibfile} 

\bsp	
\label{lastpage}
\end{document}